\documentclass{iau} 
\usepackage{graphicx}

\title[The link between hot and cool outflows] 
{The Link between Hot and Cool Outflows}

\author[Jorick S. Vink et al.] 
{Jorick S. Vink, A.A.C. Sander, E.R. Higgins, G.N. Sabhahit}

\affiliation{Armagh Observatory and Planetarium, College Hill, BT61 9DG, Armagh, Northern Ireland \\email: {\tt jorick.vink@armagh.ac.uk}}

\pubyear{2022}
\volume{366}  %% insert here IAU Symposium No.
\jname{The Origin of Outflows in Evolved Stars}
\editors{A.C. Editor, B.D. Editor \& C.E. Editor, eds.}
\begin{document}

\maketitle

\begin{abstract}
The link between hot and cool stellar outflows is shown to be critical for correctly predicting the masses of the most massive black holes (BHs) below the so-called pair-instability supernova (PISN) mass gap. Gravitational Wave (GW) event 190521 allegedly hosted an "impossibly" heavy BH of 85\,$M_{\odot}$. Here we show how our increased knowledge of both metallicity $Z$ {\it and} temperature dependent mass loss is critical for our evolutionary scenario of a low-$Z$ blue supergiant (BSG) progenitor of an initially approx 100\,$M_{\odot}$ star to work. We show using MESA stellar evolution modelling experiments that as long as we can keep such stars above 8000 K 
such low-$Z$ BSGs can avoid strong winds, and keep a very large envelope mass intact before core collapse. This naturally leads to the Cosmic Time dependent maximum BH function below the PISN gap.
\end{abstract}

%\firstsection % if your document starts with a section,
              % remove some space above using this command.

\section{Introduction}

Accurate mass-loss rates -- as a function of effective temperature $\dot{M} = f(T_{\rm eff})$ --
are needed for making reliable predictions for the evolution of the most massive stars, including
the black hole (BH) mass function with respect to metallicity $Z$ (see Sander et al. these proceedings).

Over the last five years gravitational wave (GW) observatories have shown the existence
of very heavy black holes. The current record holder is the primary object in the GW event 190521 with 85\,$M_{\odot}$. 
Due to the fact that this BH mass is almost twice as large as the generally accepted lower boundary of the pair-instability (PI) supernova (SN) mass gap at approximately 50\,$M_{\odot}$ (Farmer et al. 2019; Woosley \& Heger 2021), the GW event discoverers argued the 85$M_{\odot}$ BH is most likely a second generation BH, as it would be {\it impossible} for a progenitor star to have directly collapsed into a BH within the PISN mass gap spanning
50 and 130 $M_{\odot}$ (Abbott et al. 2020).

In this contribution we show that this conclusion could be premature, as we have constructed a robust blue supergiant (BSG) scenario
for the collapse of a very massive star (VMS) of order 100\,$M_{\odot}$ at low $Z$ (Vink et al. 2021). The key physiscs involves both 
the $Z$-dependence and the effective temperature dependence of the mass-loss rate of evolved supergiants. 

\begin{figure}
\begin{center}
 \includegraphics[width=5.4in]{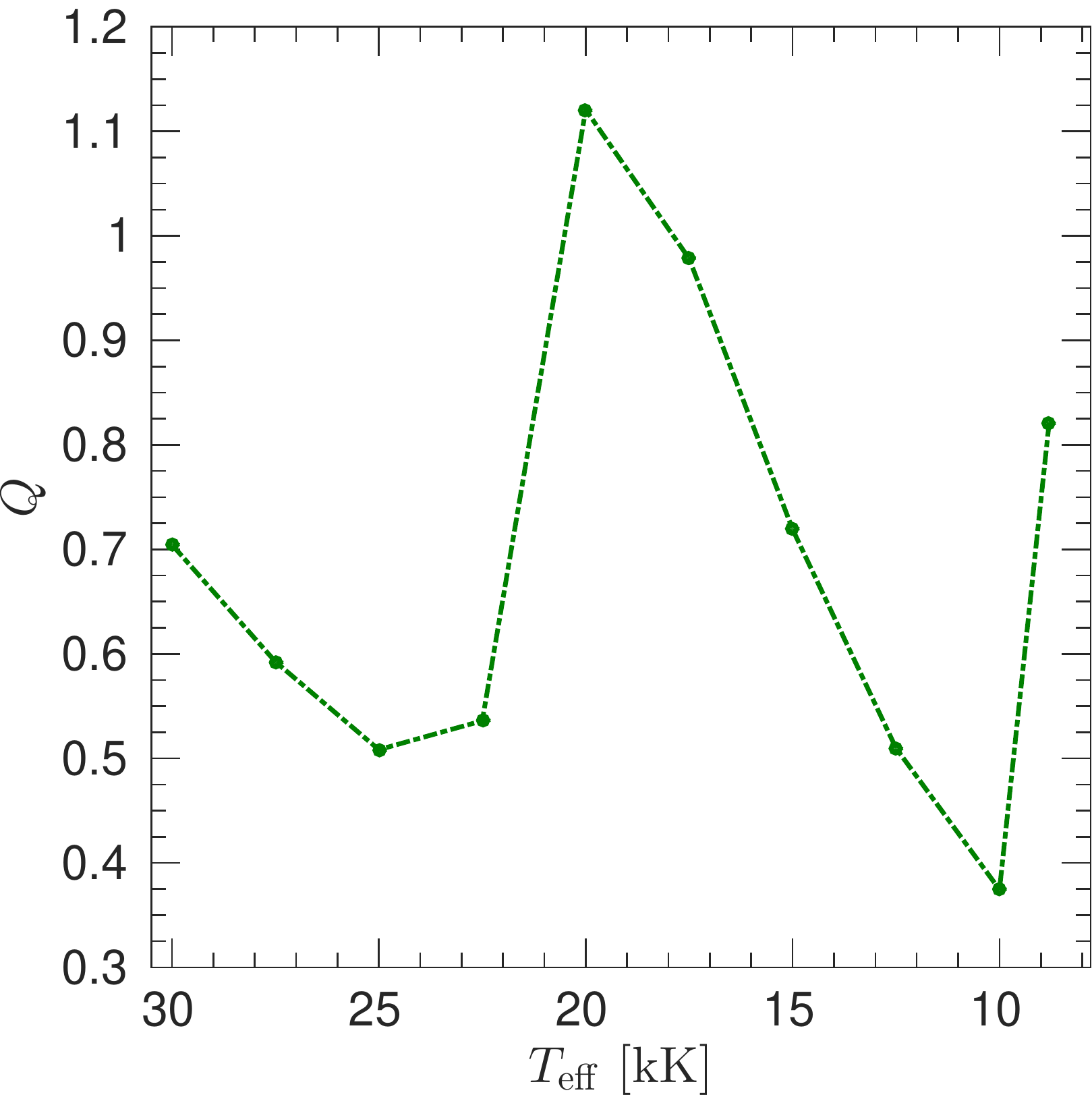} 
 \caption{The first and second bi-stability jump in terms of the global wind parameter $Q$ (see Petrov et al. 2016.). Note that the exact location of the second jump is below the lower temperature boundary, i.e. with $T_{\rm eff}$ lower than 8800 K.}
   \label{fig1}
\end{center}
\end{figure}

\section{Overview of hot and cool mass-loss rates}

When a massive star burns hydrogen (H) in the core it traditionally evolves from the hot blue side to the cool 
red part of the stellar Hertzsprung-Russell diagram (HRD).
When this takes place at approximately constant luminosity $L$ the key physical parameters are (i) the amount of mixing by processes such as core overshooting,
as these set the duration of the wind mass-loss phase during H-burning and the total mass being lost on the main sequence, as well as (ii) the absolute rate of mass-loss (dependent on the host galaxy $Z$) and (iii) how this mass loss varies from the hot to the cool side of the HRD.
For hot stars above 10 kK the winds are driven by gas opacity (see Vink 2022 for a recent review) and while the absolute mass-loss values are
still under debate (e.g. Bj\"orklund et al. 2021) the implication that mass-loss rates of hot-star winds are $Z$-dependent is undisputed. The exact $Z$ dependence still needs to be established however (Vink \& Sander 2021).

When $T_{\rm eff}$ drops during stellar evolution -- starting from approx. 40 kK -- the mass-loss rate is first expected to drop (see Fig.\,1).
The reason for this behaviour is that the line acceleration is set by the product of the stellar flux and the opacity.
When $T_{\rm eff}$ drops, the stellar flux gradually moves from the ultraviolet (UV) part of the spectral energy distribution (SED) to the optical,
while the opacity is still predominately 'left behind' in the UV part of the SED. In other words,
there is a growing mismatch between the flux and the opacity, implying the flux-weighted opacity drops, 
and so does $\dot{M}$ (Vink et al. 1999).

This situation changes abruptly when the dominant line driving element iron (Fe) recombines, causing a bi-stability jump (BSJ) in the wind parameters. The first recombination is that from Fe {\sc iv} to {\sc iii} at approx. 21 kK, the second recombination from Fe {\sc iii} to {\sc ii} takes place below 8800 K (Vink et al. 1999; Petrov et al. 2016). The exact location of this second BSJ has not yet been determined, as the current generation of sophisticated co-moving frame (CMF) radiative transfer model atmospheres has not yet been able to converge below the recombination of H at approx 8000 K. This uncertainty of mass-loss in the yellow supergiant / hypergiant phase is of key relevance
for setting the Humphreys-Davidson (HD) limit (Gilkis et al. 2021; Sabhahit et al. 2021) and YSG mass-loss should therefore play an important role  empirically (e.g. Koumpia et al. 2020; Oudmaijer \& Koumpia these proceedings). Accurate mass-loss rates of $T_{\rm eff}$ dependent $\dot{M}$ are also critical for constructing the next generation of hydrodynamical stellar evolution models for both luminous blue variable and YSG phases 
(Grassitelli et al. 2021).

\section{Current mass-loss recipes in use - and the link to wind physics}

\begin{figure}
\begin{center}
 \includegraphics[width=5.4in]{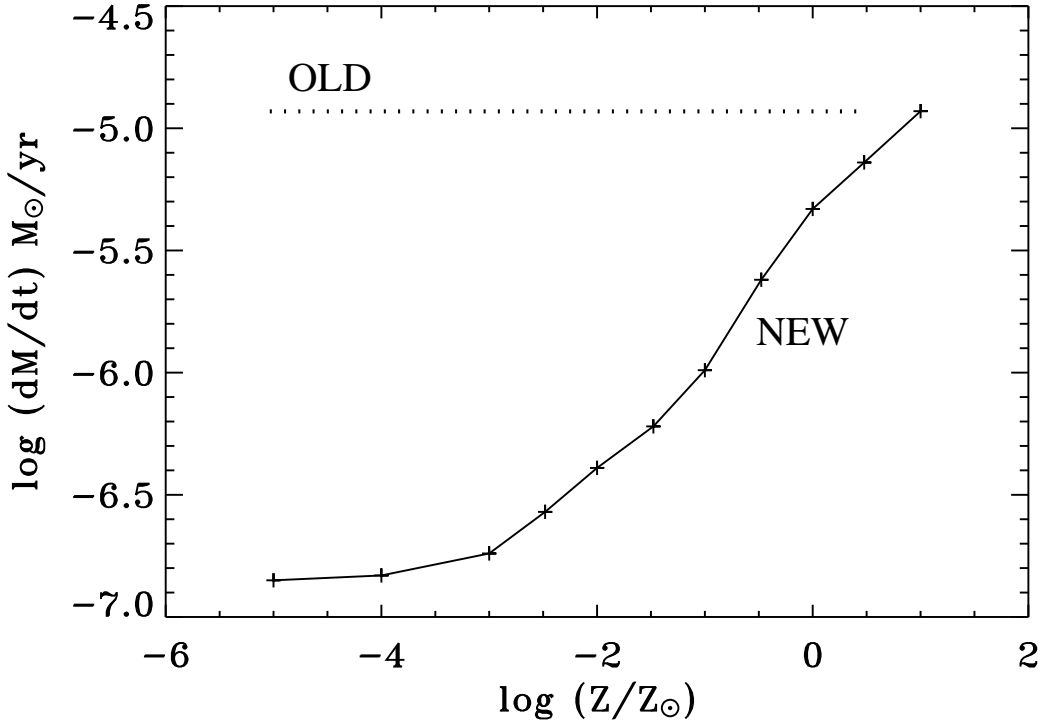} 
 \caption{The $Z$ dependence of WR stars. In stellar models before 2005 the mass-loss rate of WR stars was assumed to be independent of the host galaxy $Z$ ("OLD") as the high abundance of self-enriched elements such as carbon (C) was thought to be dominant. Vink \& de Koter (2005) showed that despite the huge amount of C in WR atmospheres the winds are nonetheless predominately driven by Fe and thus strongly Fe-dependent, indicated by "NEW". See Sander et al. these proceedings for new mass-loss predictions.}
   \label{fig2}
\end{center}
\end{figure}

One of the most used mass-loss recipes currently in use in massive star models is the "Dutch" wind loss recipe in MESA (Paxton et al. 2013).
In this collection of mass-loss prescriptions, massive stars undergo modest mass loss on the main-sequence and enhanced mass loss below the first BSJ according to Vink et al. (1999, 2001). The second BSJ is not directly covered in the Dutch recipe, but it follows a similar approach as Brott et al. (2011) where the second BSJ is {\it in}directly included by switching from the Vink et al. theoretical recipe to the empirical cool star recipe of de Jager et al. (1988) at approx. 10 kK.
This prescription yields relatively large mass-loss rates, although the physics of cool red supergiant (RSG) winds is still under debate, and this also means that whether or not RSG winds have a $Z$-dependence is presently unclear.

\begin{figure}
% \vspace*{-2.0 cm}
\begin{center}
 \includegraphics[width=5.4in]{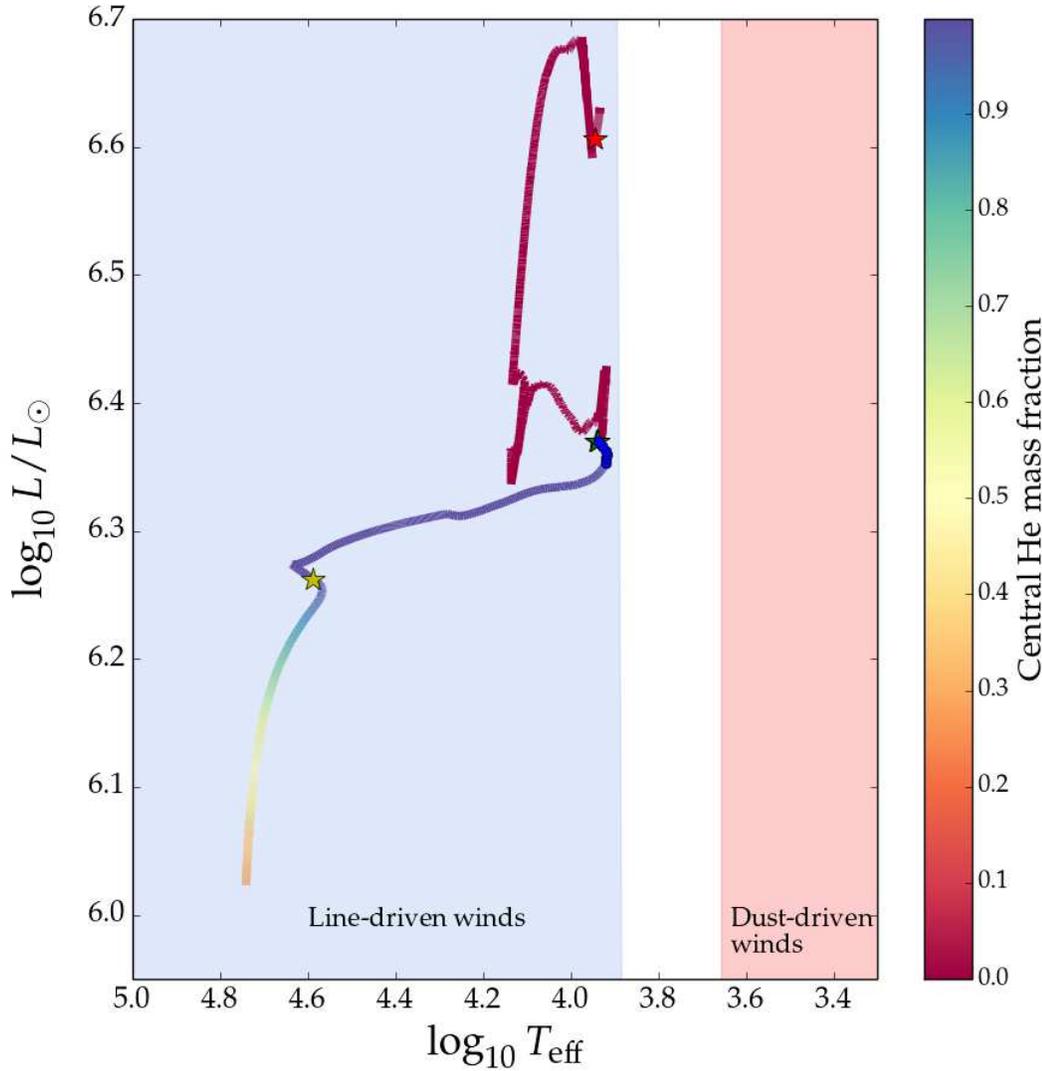} 
% \vspace*{-1.0 cm}
 \caption{Evolution of our BSG progenitor model in a Hertzsprung-Russell diagram (HRD). 
The colour bar represents the core He abundance, with a yellow star showing the TAMS position, a blue star illustrating the end of core He-burning, and a red star marking the end of core O-burning. Blue dots (near the blue star) show time-steps of 50,000 years after core H exhaustion, where time is spent as a BSG (i.e. above 8 kK). Shaded regions highlight the area in the HRD where RSGs (red) evolve with dust-driven winds (as generally assumed, but the physical mechanism is still debated) or BSGs (blue) evolve with line-driven winds.}
   \label{fig3}
\end{center}
\end{figure}

For stars evolving back to the hotter part of the HRD above 10 kK and with enriched atmospheres with a helium mass fraction 
$Y$ larger than approx. 0.6, the models generally 
assume the total $Z$ -- including all elements heavier than He -- empirical Wolf-Rayet (WR) recipe of 
Nugis \& Lamers (2000). Note that a scaling with this 'total' $Z$ is unlikely to be physically correct, as Vink \& de Koter (2005) showed the host galaxy Fe to be the main wind driver despite the larger abundances of self-enriched elements (see Fig.\,2).

In order to account for Fe-dependent winds as well as the knowledge that the second BSJ is located below 10 kK and even below 8800 K, Vink et al. (2021) and Sabhahit et al. (2021) provided an updated version of the Dutch wind mass-loss recipe in MESA. With this improved treatment, stars typically have lower mass-loss rates in the BSG phase, which prevents excessive mass loss at low $Z$, and helps stars maintain sufficient envelope mass to form very heavy BHs as long as they remain hotter than $\sim$8000 K (see Fig.\,3).

\section{Implications for impossible black holes over Cosmic Time}

\begin{figure}
% \vspace*{-2.0 cm}
\begin{center}
 \includegraphics[width=5.4in]{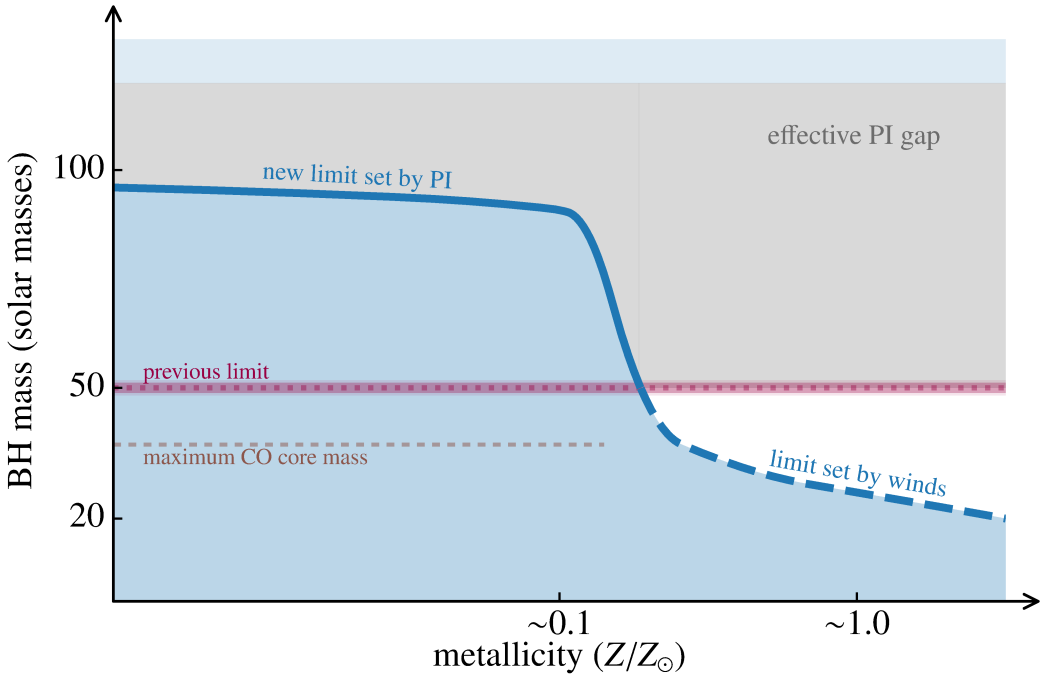} 
% \vspace*{-1.0 cm}
 \caption{Maximum black hole mass as a function of $Z$ or Cosmic Time. At low $Z$ the maximum mass from our models is effectively doubled in comparison to earlier models and assumptions, while the maximum black hole mass at higher $Z$ is set by stellar winds.}
   \label{fig4}
\end{center}
\end{figure}

Vink et al. (2021) realized that in order to enable the collapse of a VMS to an 85\,$M_{\odot}$ BH this needs 2 key ingredients. The first one is an intrinsically low $Z$ in order for $Z$-dependent mass loss to not evaporate the initial stellar mass. The second requirement involves a relatively low amount of core overshooting, equivalent to a step overshooting parameter $\alpha_{\rm ov}$ of 0.1 or below. The reason for this second ingredient is 3-fold.

Firstly, low overshooting keeps the star more compact, and the collapse of the entire envelope of a very massive BSG is easier to accomplish than that for a RSG (e.g. Fernandez et al. 2018). Secondly, a low overshooting keeps the core mass below the PISN limit, and enforces a larger envelope mass (e.g. Higgins \& Vink 2019).
Thirdly, if the star remains above the effective temperature of the second BSJ, the regime of high mass-loss rates at low $T_{\rm eff}$ can be avoided (see Figs.\,1 \& 3).

  Vink et al. (2021) showed that at $Z$ values below approx. 10\% of the solar metallicity 90-100 $M_{\odot}$ stars initially could have core masses below the critical 37\,$M_{\odot}$ limit, and collapse into 80-90 $M_{\odot}$ BHs. Such impossibly heavy BHs are firmly within the canonical PISN mass gap, which should therefore be adjusted.
A schematic maximum BH mass with $Z$ is shown in Fig.\,4. The figure shows an almost twice as large maximum upper BH mass below the PISN gap at early Cosmic times at low $Z$, while for larger $Z$ the maximum BH mass is directly set by stellar wind mass loss.

\end{document}